\documentclass[12pt,a4paper]{article}
\usepackage{amsmath,amsxtra,amsfonts,amssymb,amstext}
\usepackage[T1]{fontenc}
\usepackage[tight]{subfigure}
\usepackage{aeguill}
\usepackage{fleqn}
\usepackage{graphicx}
\usepackage{enumerate}
\usepackage{bm,here}
\usepackage{mathptm}

\newcommand{\ppt}[1]{\frac{\partial{#1}}{\partial t}}
\renewcommand{\vec}[1]{\ensuremath\boldsymbol{\mathrm{#1}}}

\newcommand{\nablab}{\boldsymbol{\nabla}}
\newcommand{\cdotb}{\boldsymbol{\cdot}}

\newcommand{\timesb}{\boldsymbol{\times}}

\def \text{\mbox}

\def\eps{\epsilon}
 
\def\grapprox{\mathop{\lower.5ex \hbox{$\buildrel{\fivesy >}\over{\fivesy\sim}$}} \nolimits}
\def\lsapprox{\mathop{\lower.5ex \hbox{$\buildrel{\fivesy <}\over{\fivesy\sim}$}} \nolimits}
\def\grls{\mathop{\lower.5ex \hbox{$\buildrel{\fivesy >}\over{\fivesy <}$}} \nolimits}

\def\vec#1{{\bf #1}}

\def\ppt#1{\partial #1/\partial t}

\def\Psi{A_\parallel}

\begin{document}

%
\title{ Impurity Transport in Plasma Edge Turbulence}
%
\author{Volker Naulin,Martin Priego Wood,  and  Jens Juul Rasmussen
\\Association EURATOM - Ris{\o} National Laboratory\\
  Optics and Plasma Research, OPL - 128\\
  DK - 4000 Roskilde, Denmark  }
\maketitle
The turbulent transport of  minority species/impurities is
investigated in  2D drift-wave turbulence as well as in  3D toroidal drift-Alfv\'en
edge  turbulence. 
The full effects of perpendicular and -- in 3D -- parallel advection
are kept for the impurity species. Anomalous pinch effects are
recovered and explained in terms of Turbulent EquiPartition (TEP)

\section{Anomalous Pinch in 2D Drift-Wave Turbulence}

The Hasegawa-Wakatani model \cite{Hasegawa:Wakatani:1983}  for 2D resistive drift-wave turbulence reads
\begin{equation}
\begin{subequations}\label{eq:hw_r}
\label{eq:hw_w_r} d_{t}{(n-x)}  = \mathcal{C} (\varphi-n) + \mu_n \nabla_\perp^2 n,\hspace{2ex}
\label{eq:hw_n_r} d_{t}{\omega} = \mathcal{C} (\varphi-n) + \mu_\omega \nabla_\perp^2 \omega,
\end{subequations}
\end{equation}
with
$\omega \equiv \nabla_\perp^2 \varphi $ and
$d_{t} \equiv \ppt{}+\vec{\hat z}\timesb\nablab_\perp\varphi \cdotb\nablab_\perp$.
Here, $n$ and $\varphi$ denote  fluctuations in density and
electrostatic potential. $\omega$ is the vorticity, 
$\nabla \times \vec{\hat z}\timesb\nablab_\perp\varphi$. The
parameters in the HW system are
the parallel coupling $\mathcal{C}$, and diffusivities, $\mu_n$,$\mu_\omega$.
\\
2D impurity transport in magnetized plasma is modeled by the transport
of a passive scalar field:
\begin{equation}
\label{eq:imp_mod_r}
d_{t}{\theta} - \zeta\nablab_\perp\cdotb\bigg(\theta d_{t}{\nablab_\perp\varphi}\bigg)=
\mu_\theta\nabla_\perp^2\theta,
\end{equation}
where $\theta$ is the density of impurities, $\mu_\theta$ the collisional diffusivity, and 
$
\zeta=\frac{m_\theta}{q_\theta}\frac{e}{m_i}\frac{\rho_s}{L_n}
$
the influence of inertia, which enters via the polarization drift. 
The latter makes the flow compressible, consequently for ideal (massless) impurities, $\zeta=0$ and 
advection is due to the incompressible electric drift only.
In all cases the impurity density is assumed to be so low compared to
the bulk plasma density that there is no back-reaction on the bulk
plasma dynamics.

\subsection{ Vorticity - Impurity correlation}
\begin{figure}
\begin{minipage}{0.450\textwidth}
\includegraphics[width=\textwidth]{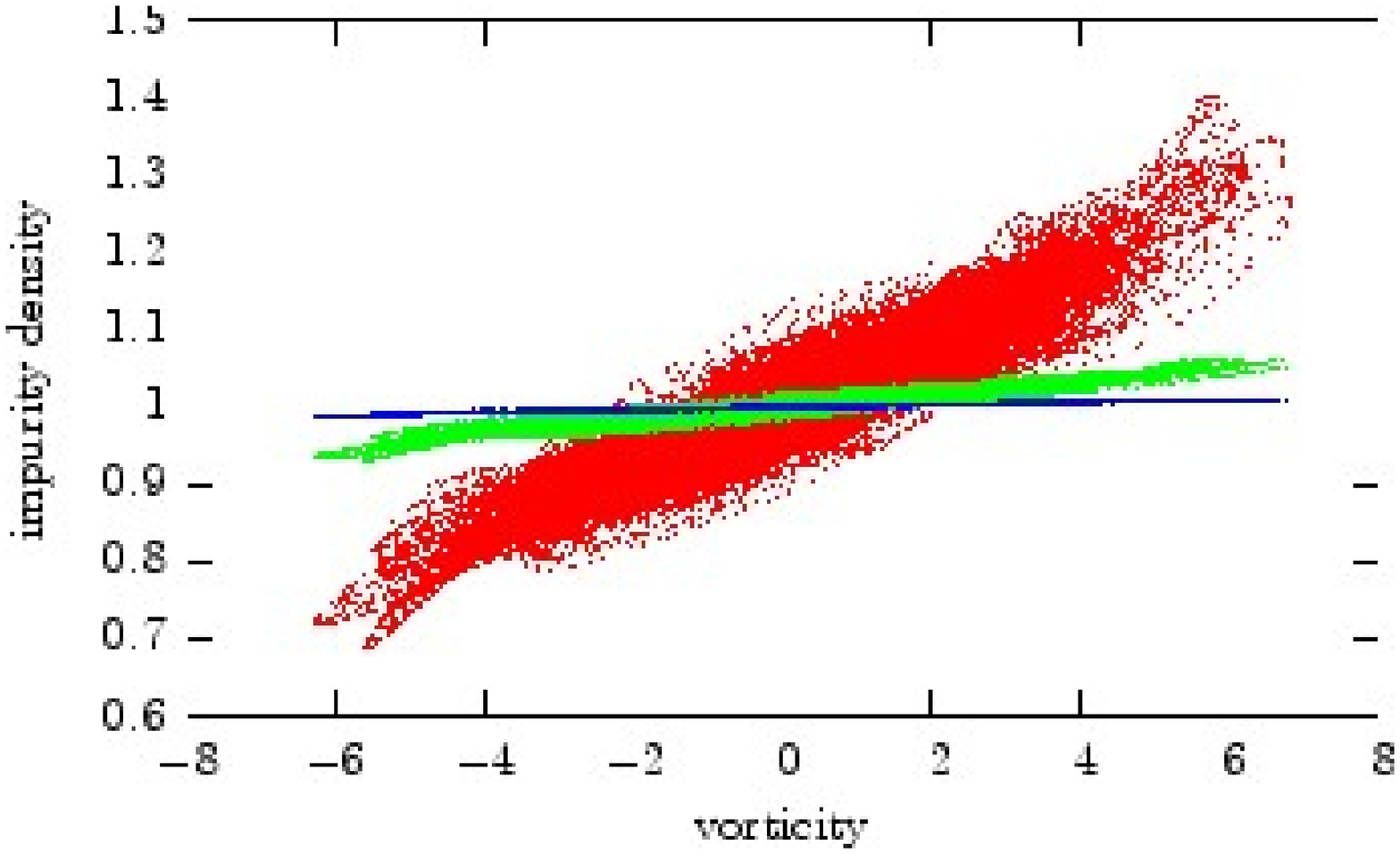}
\caption{Scatter plot of impurity density and the vorticity field at $t=100$
for different values of the mass--charge ratio $\zeta$ in the saturated state in
HW with $\mathcal{C}=1$: $\zeta=0.05$ (red), $\zeta=0.01$ (green), and
$\zeta=0.002$ (blue). }
\label{fig:hw1cp_vp}
\end{minipage}
\hfill
\begin{minipage}{0.45\textwidth}
\includegraphics[width=\textwidth]{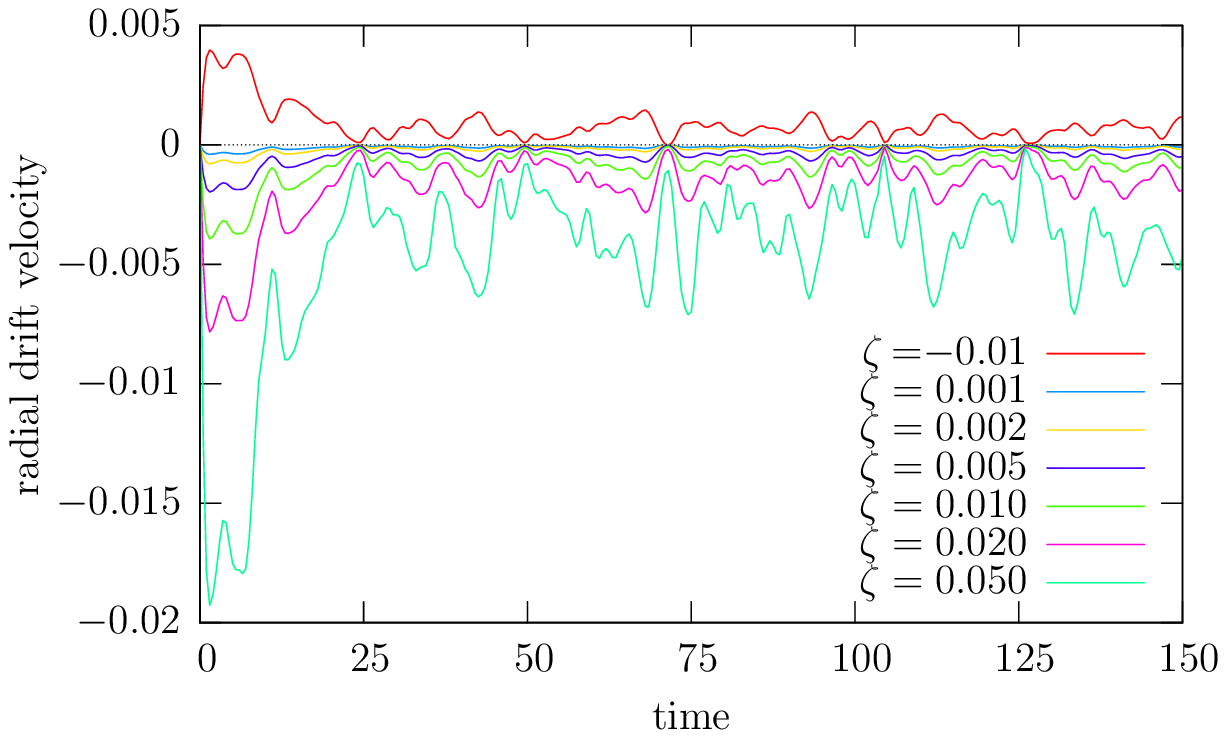}
\caption{Evolution of the radial drift velocity of inertial impurities in the
  saturated state in HW with $\mathcal{C}=1$. The impurities are uniformly
  distributed at $t=0$.}
\label{fig:hw1cp_xdrift}
\end{minipage}
\end{figure}

The equation for the impurities can be rewritten in the form: 
\begin{equation*}
d_{t} {(\ln\theta-\zeta\omega)}
=\zeta\nablab_\perp\ln\theta \cdotb
d_{t} {\nablab_\perp\varphi} + \frac{\mu_\theta}{\theta}\,\nabla_\perp^2\theta\;.
\end{equation*}
If the
diffusivity $\mu_\theta$ is of order $\zeta \ll 1$ and fluctuations
$\theta_1 $ of the impurity density  measured relative to a constant
impurity background $\theta_0$ do not exceed a corresponding level, the quantity
$\ln\theta-\zeta\omega$ is approximately a Lagrangian invariant.
Turbulent mixing will homogenize Lagrangian invariants in 
TEP states \cite{Yankov:1997,Naulin:Rasmussen:Nycander:1998}, leading to 
\begin{equation*}
\ln\theta-\zeta\omega\approx\mathrm{const},
\end{equation*}
which constitutes a prediction about the effect of compressibility on the
initially homogeneous impurity density field. 
The conservation of  impurity density yields 
\begin{equation*}
\frac{\theta}{\theta_0}\approx 1+\zeta\omega,
\end{equation*}
which conforms with the assumed ordering.
We thus predict a linear
relation between  impurity density $\theta$ and  vorticity $\omega$, the
proportionality constant being the mass--charge ratio $\zeta$. 
This is related, but not the same as, 
to the aggregation of dense particles in vortices in 
fluids due to the Coriolis force \cite{Bracco:etal:1999}.
\begin{figure}
  \centering
  \subfigure[][]{
    \includegraphics[width=.45\textwidth]{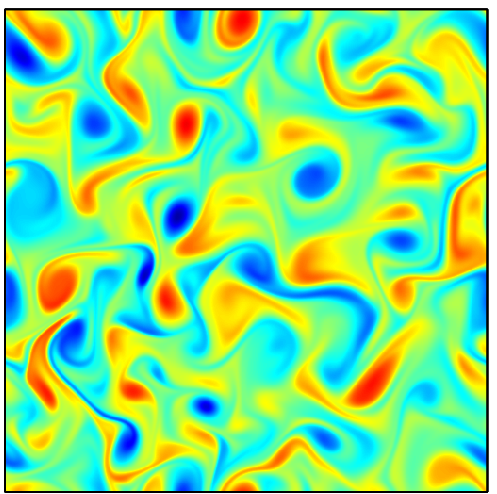}}
  \hfill
  \subfigure[][]{
    \includegraphics[width=.45\textwidth]{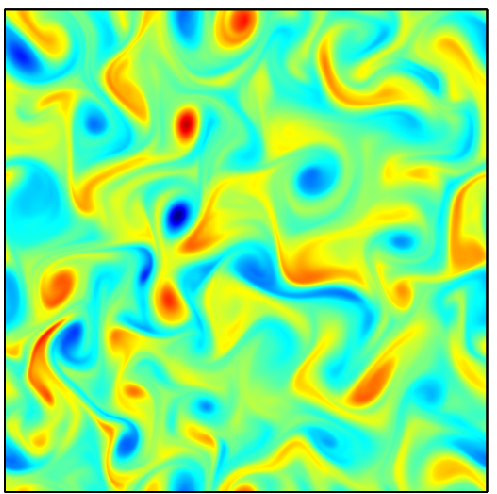}}
\caption{(a) Vorticity  and (b) density  of inertial impurities in the
  saturated state with $\mathcal{C}=1$ and
  $\zeta=0.01$, L = 40. Other 
parameters:
  $\mu_n=\mu_\omega=\mu_\theta=0.02$.}
\label{fig:hw1cp3}
\end{figure}
The prediction is verified by numerical simulations 
 of inertial impurities in saturated HW-turbulence  for
 $\mathcal{C}=1$. 
The simulations are performed on a $[-20,20]^2$
domain, using $512^2$ gridpoints, and impurity
diffusivity $0.02$. The impurity density field is initially 
set to unity. The impurity density field for $\zeta=0.01$
is presented together with  vorticity in Figure~\ref{fig:hw1cp3}.
 Figure~\ref{fig:hw1cp_vp} shows a
scatter plot of the point values of impurity density and vorticity at time $150$
for three different values of $\zeta$.  The proportionality factor
$\theta = 1 + K \omega $ is
determined to be slightly below one: $K\simeq0.82\,\zeta$. 

\subsection{Anomalous pinch}
The role of inertia for a radially inward
pinch is investigated by considering the
collective drift of impurities.
Ideal impurities do on average not experience a drift, but 
this is not the case for
inertial impurities, since compressibility  effects arrange for a
correlation between $\theta_1$ and $\omega$. Note that only the
deviations from the above discussed linear relationship $\theta = 1 +
K \omega $ result in a net flow, as $\int K \omega v_r \,dx = 0 $ for
periodic boundary conditions.

The evolution of the radial
drift velocity, measured as the net radial impurity transport,  is
presented in  
Figure~\ref{fig:hw1cp_xdrift}. The radial
drift velocity has a definite sign that depends on the sign of $\zeta$. There is a
continuous flow of impurities in a definite direction (inward for positively
charged impurities). This  resembles the anomalous pinch observed in
magnetic confinement experiments \cite{Dux:2003}.  
 Average radial drift velocities computed using the values of the drift
from $t=25$ to $t=150$ are presented in Table~\ref{tab:xdrift}. The scaling of the average radial
drift with $\zeta$ is seen to be remarkably linear.
\begin{table}
\centering
\caption{Radial drift velocity of impurities for different values of the
mass--charge ratio $\zeta$ in the saturated state in HW with $\mathcal{C}=1$.
Calculated as the average value between $t=25$ and $t=150$. 
Parameters: $\mu_n=\mu_\omega=\mu_\theta=0.02$.}
\label{tab:xdrift}
\begin{minipage}[c]{0.30\textwidth}
\begin{tabular}{r@{\hspace{0.5cm}}c}
\hline
\multicolumn{1}{c}{$\zeta$} & radial drift\\
\hline
$-0.010$ & $\phantom{-}8.67\times10^{-4}$\\
$0.001$  & $-8.66\times10^{-5}$\\
$0.002$  & $-1.73\times10^{-4}$\\
$0.005$  & $-4.35\times10^{-4}$\\
$0.010$  & $-8.69\times10^{-4}$\\
$0.020$  & $-1.75\times10^{-3}$\\
$0.050$  & $-4.55\times10^{-3}$\\
\hline
\end{tabular}
\vspace{1em}
\end{minipage}
\hfill
\begin{minipage}{0.55\textwidth}
\includegraphics{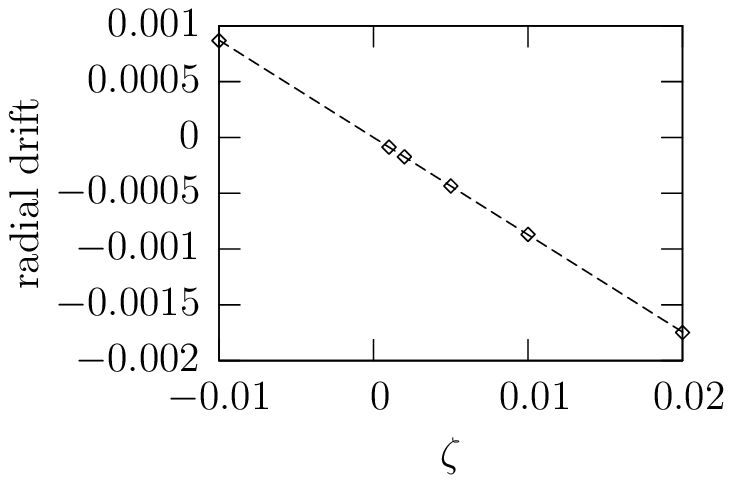}%
\end{minipage}
\end{table}
%
\section{Drift-Alfv\'en Turbulence}
\begin{figure}
\centering \includegraphics[width=0.7\textwidth]{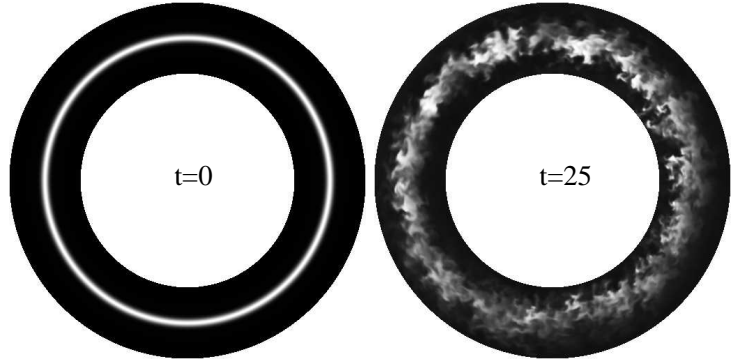}
\caption{Impurity distribution projected onto a poloidal cross-section
  (radial dimension not to scale). Initial distribution (left) and after 25 time
  units (right).  
}
\label{fig:ring}
\end{figure}

\begin{figure}
 \centering
  \subfigure[][]{
    \includegraphics[width=0.45\textwidth]{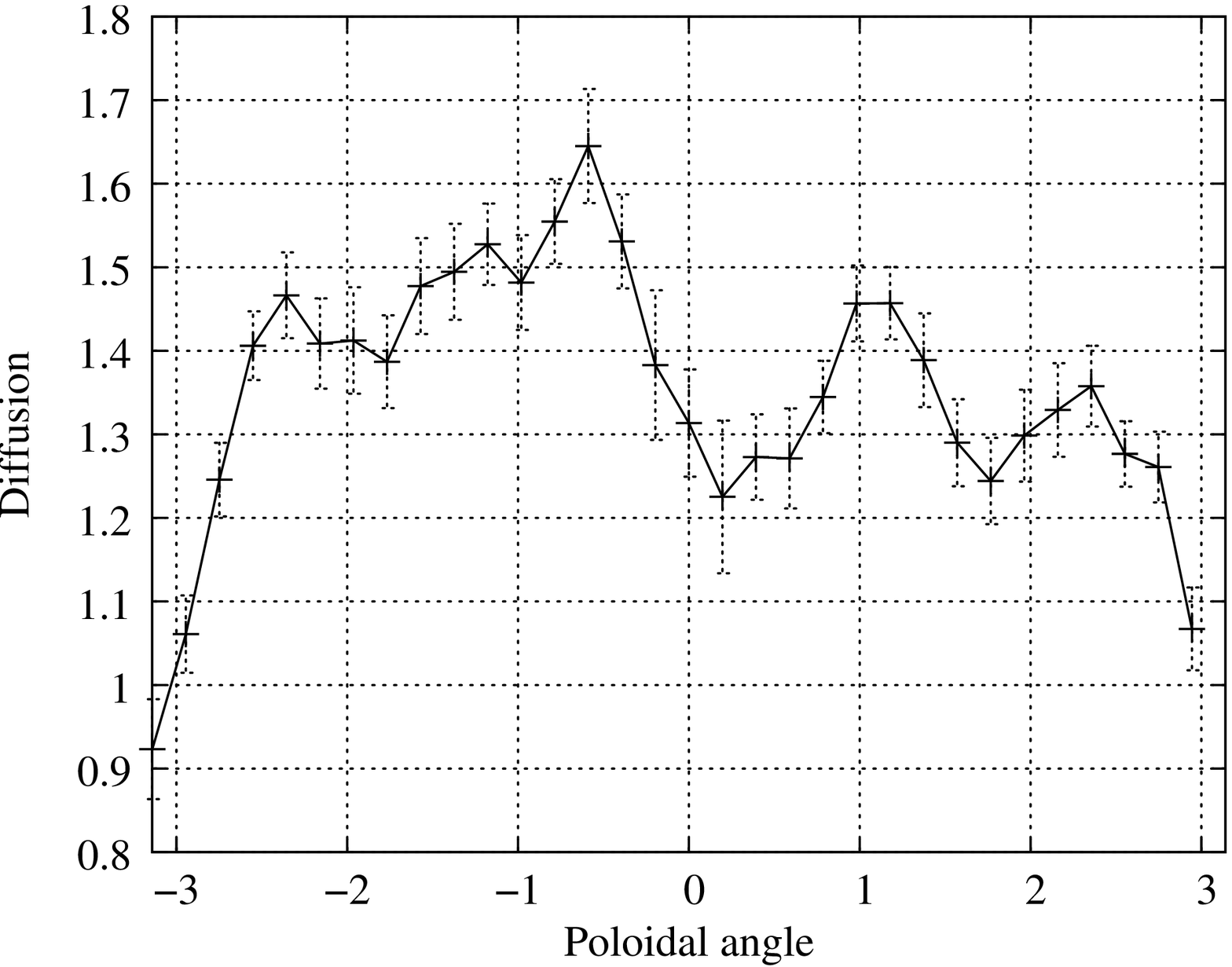}}
  \hfill
  \subfigure[][]{
    \includegraphics[width=0.45\textwidth]{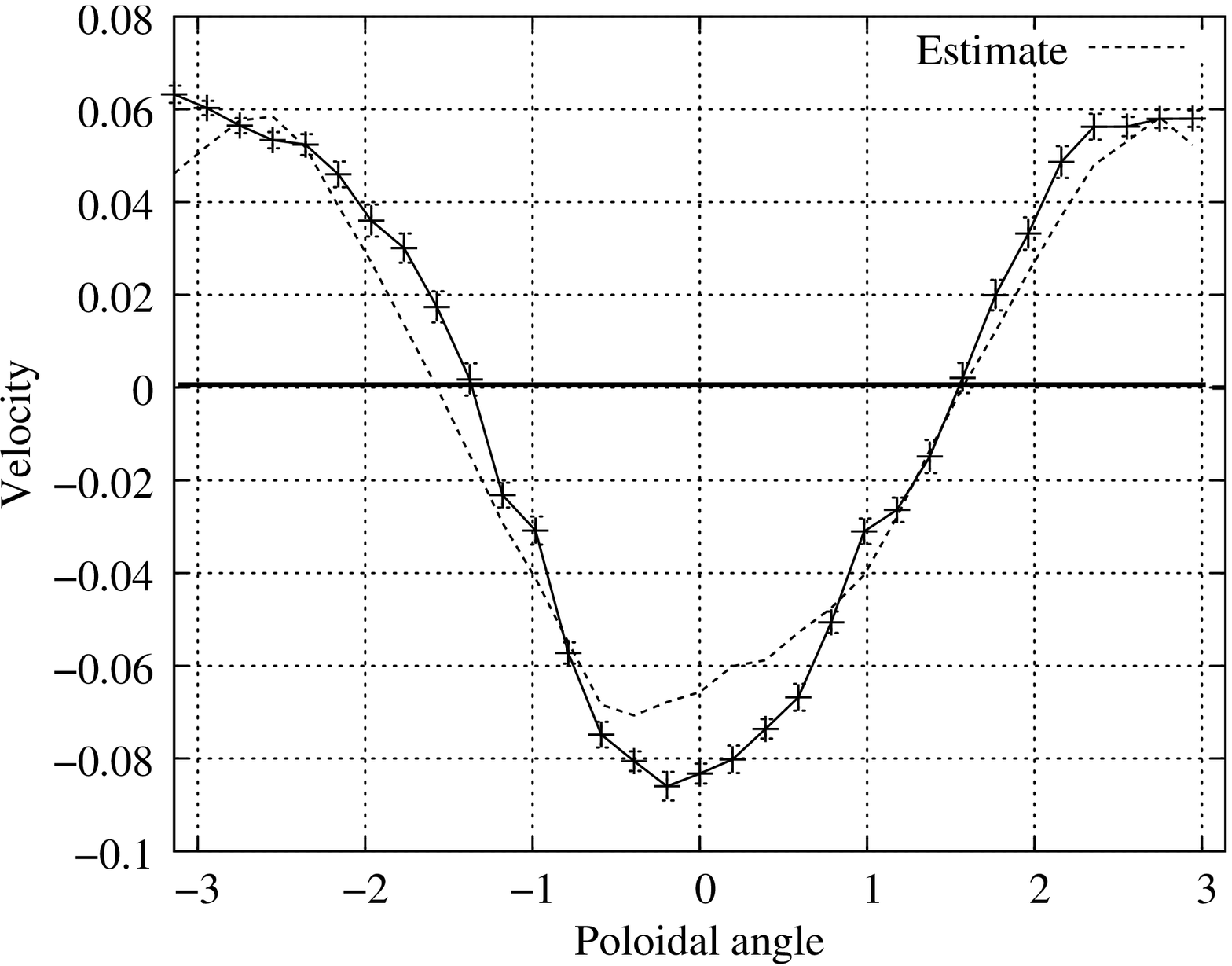}}
\caption{Impurity diffusion  $D$ (a)  and pinch velocity  $V$ (b) over
  poloidal position ($s$) with error-bars. The pinch velocity is compared to
$\omega_b*\cos(s)*D(s)$ (dashed line).}
\label{fig:vd}
\end{figure}

We now consider drift-Alfv\'en turbulence in flux tube geometry ~\cite{Scott:1997:1,Scott:1997:2,Naulin:2003}.
The following equations for the fluctuations in density $n$, potential
$\phi$ with associated vorticity $\omega = \nabla_\perp^2 \phi$, current
$J$ and parallel ion velocity $u$ arise in the usual drift-scaling:
\begin{subequations}
\begin{gather}
\frac{\partial\omega}{\partial t} + \{ \phi, \omega\} =
\mathcal{K}\left( n \right) + \nabla_\shortparallel J 
+ \mu_\omega\nabla_{\perp}^{2}\omega, \label{eq:eqvor} \\
\frac{\partial n}{\partial t} + \{ \phi, n_{EQ} + n \} = 
\mathcal{K}\left( n  - \phi \right) +  \nabla_\shortparallel
\left( J  -  u \right) + \mu_{n}\nabla_{\perp}^{2} n , \label{eq:eqne} \\
\frac{\partial}{\partial t} \left( \widehat{\beta}A_{\shortparallel} +
\widehat{\mu} J \right) + \widehat{\mu}\left\{ \phi , J \right\} = 
\nabla_\shortparallel \left( n_{EQ} + n - \phi \right) - C J , \label{eq:eqpsi} \\
\widehat{\eps}\left(\frac{\partial u}{\partial t} +
\left\{ \phi , u \right\} \right) = - \nabla_\shortparallel
\left( n_{EQ} + n \right) . \label{eq:equi} \end{gather}
\end{subequations}
The evolution of the impurity density is given by 
\begin{equation}
d_{t} \theta
=  (\zeta/\widehat{\eps}) \nabla_{\perp}\cdot \left( \theta d_{t} \nabla_{\perp} \phi  \right) 
- \theta \mathcal{K}\left( \phi \right) 
- \nabla_{\|} \left( \theta u \right) - \mu_{\theta} \nabla_{\perp}^{2} \theta
\label{eq:impurity}
\end{equation}
Standard parameters for simulation runs were $\widehat{\mu}=5$, $q = 3$,  magnetic
shear $\widehat{s}=1$, and
$\omega_{B}=0.05$, with 
$\mu_\omega=\mu_n=0.025$, corresponding to typical edge parameters of
large fusion devices. Simulations were performed on a grid with $128 \times 512
\times 32$ points and dimensions $64 \times 256 \times 2 \pi$ in
${x,y,s}$ corresponding to a typical approximate dimensional size of 2.5 cm
$\times$ 10 cm$\times$ 30 m  \cite{Scott:1997:1}.
Here we present results from a low $\hat \beta =  0.1$ run with $C =
11.5$. In Figure \ref{fig:ring} the dynamical evolution of the impurity density
is exemplified in a plot showing the poloidal projection of the
impurity density. 
\\
The
flux $\Gamma$ of the impurity ion species 
can in lowest order be expressed by the standard parameters
used in  modeling and in evaluation of transport experiments: a diffusion coefficient $D$ and a
velocity $V$, which is associated to a pinch effect,
\begin{equation}
\Gamma_{y} (s) = -D(s) \partial_{x} \left<\theta\right>_{y} + V(s)
\left< \theta \right>_{y}\,.
\label{eq:dv}
\end{equation}
From scatter plots of  $ \Gamma(r)/\left<n\right >_{y}  $ versus $\partial_x
\ln \left< n\right>_{y}$, values for   $D(s)$
and $V(s)$ are obtained.  
The poloidal (coordinate $s$) dependence of 
$D$ and $V$ is rather strong and
shown, with numerical uncertainties,  in
Fig.~\ref{fig:vd}.
The effective advective  velocity $V(s)$ changes sign and is at the high
field side directed outwards.  This pinching velocity is due to 
curvature and can be consistently explained in the framework of Turbulent
EquiPartition (TEP)
\cite{Nycander:Yankov:1995,Naulin:Rasmussen:Nycander:1998} as follows:
In the absence  of parallel advection, finite
mass effects and diffusion,  Eq.~(\ref{eq:impurity})  has the following
approximate Lagrangian invariant
\begin{equation}
L(s) = \ln \theta + \omega_B x \cos(s)  - \omega_B y \sin(s) \;.
\label{eq:tep}
 \end{equation}
TEP assumes the spatial  homogenization of $L$ by the
turbulence. As parallel transport is weak,  each drift plane
$s=\text{const.}$  homogenizes independently. This leads to profiles
$\left<L(s) \right>_y
= \text{const.}(s)$.
At the
outboard midplane ($s=0$) the impurites are effectively advected
radially inward leading to an impurity profile  ($ \left< \ln  \theta\right>_y
\propto  const.   -\omega_B  x   $), while at the high field side they are
effectively advected outward ($ \left< \ln \theta \right>_y
\propto   const. +  \omega_B  x   $).
One should note that this effective inward or outward advection
is not found as an average $E \times B$ velocity, but is mitigated
by the effect of spatial homogenization of $L$  
under the action of the turbulence. 
The strength of
the ``pinch'' effect is consequently proportional to the mixing properties of the
turbulence and  scales
with the measured effective turbulent diffusivity. We arrive at the following
expression for the connection between pinch and diffusion: 
\begin{equation}
V(s) =  - \alpha \omega_B \cos (s) D(s)\; .
\end{equation}
Considering a stationary case with zero flux and Eq.~(\ref{eq:tep})  
we obtain $\alpha = 1$. 
The  ballooning in the
turbulence level  causes  the inward flow  on the outboard midplane to
be stronger
than the effective outflow on the high-field side. Therefore, averaged over a flux surface
and assuming a poloidally constant impurity density,  a net impurity
inflow results. This net pinch is proportional to the diffusion
coefficient $D$ in agreement with experimental observations
\cite{Perry:Brooks:Content:etal:1991}. 

{\bf Acknowledgement:} Extensive discussions with O.E. Garcia are
  gratefully acknowledged.


\end{document}